\documentclass[12pt]{article}
\usepackage{graphicx}
\begin{document}

\begin{titlepage}
\begin{center}
{\Large \bf Recurrence metrics and time varying light cones }

\vspace{5mm}

\end{center}

\vspace{5 mm}

\begin{center}
{\bf Moninder Singh Modgil \footnote[1]{ Department of Physics,
Indian Institute of Technology, Kanpur, India,\\email:
msingh@iitk.ac.in, moni\_g4@yahoo.com} }

\vspace{3mm}

\end{center}

\vspace{1cm}

\begin{center}
{\bf Abstract}
\end{center}
It is shown by explicit construction of new metrics, that General
Relativity can solve the exact Poinc$\acute{a}$re recurrence
problem. In these solutions, the light cone, flips periodically
between past and future, due to a periodically alternating arrow
of the proper time. The geodesics in these universes show periodic
Loschmidt's velocity reversion $v \rightarrow -v$, at critical
points, which leads to recurrence. However, the matter tensors of
some of these solutions exhibit unusual properties - such as,
periodic variations in density and pressure. While this is to be
expected in periodic models, the physical basis for such a
variation is not clear. Present paper therefore can be regarded as
an extension of Tipler's "no go theorem for recurrence in an
expanding universe", to other space-time geometries.

\vspace{1cm} {\noindent \bf KEY  WORDS:} General Relativity,
Poinc$\acute{a}$re Recurrence, Black holes, White holes, Closed
Time-like Curves.

\end{titlepage}

\section{Introduction}
The concept of recurrence metrics, introduced in this paper
derives its motivation from the Poinc$\acute{a}$re recurrence
problem, i.e., – given a system of $N$, particles, under what
conditions, will the system return to its initial configuration in
the phase space? In an unbounded flat Newtonian space-time,
non-interacting particles move on straight lines, and any initial
configuration would never recur. Recurrence is also inconceivable
in interacting case in a Newtonian universe, given the usual
physical potentials for particle interactions. Curved space-times,
with closed geodesics, however offer another approach to the
Poinc$\acute{a}$re recurrence problem. On a space such as $S^n$,
or $P^n$, any set of non-interacting particles, all having an
identical uniform velocity $v$, the initial configuration would
recur after a period $T=vC$, where $C$ is the circumference of
these closed ($S^n$ or $P^n$) spaces. $S^n$ and $P^n$ are termed
as example of Zoll phenomena \cite{Zoll_1903}, i.e. closure of all
geodesics in an identical period. A Lorentzian manifold, all of
whose null geodesics are closed, is said to have a Zollfrei metric
[2]. Examples of Zollfrei metric are $S^n \times S^1$ and $P^n
\times S^1$ space-times. Here $S^1$ is the time topology, and
$S^n$ and $P^n$ are topologies of spatial sections. These models
are cyclic in the sense, that each event has an infinite number of
copies in past and future. A set of non-interacting photons in
such universes, would circle the closed universe in a single
cycle, and the initial configuration would recur. Such compact
space-times offer freedom from infra-red divergences of quantum
field theory \cite{Schecter_1977, Villaroel_1986}. $S^3 \times
S^1$ is also the basis of Segal's \cite{Segal_1976} cosmological
model. However, particles slower than light would take longer to
circum-navigate the closed universe. An initial configuration of
particles, with different initial velocities, e.g. a Boltzmann
distribution, would therefore not recur. Hence, we are lead to the
question, whether there exist general relativistic metrics, where
recurrence occurs for a wider set of initial data, i.e., the
complete range of particle velocities. It would also be desirable
if this occurs for interacting case as well. as free particles
move along geodesics, therefore, for non-interacting particles, we
require equi-period closure all geodesics, for recurrence. In
interacting case - assuming the interactions to be point-like, the
particles will be switching from one geodesic to another. The
particle paths therefore, can be described as piece wise geodesic.
In this paper we present simple solutions of Einstein's field
equations, where the recurrence of initial configuration occurs,
due to equi-period closure of geodesics..

While spatial variation of light cone semi angle is well know, in
general relativity, e.g., the Schwarzschild, and G\"{o}del
\cite{Godel_1949} metrics, but periodic, temporal variation of
light cone semi angle has not received much attention. These
solutions depict, a non-uniform flow of time, both temporally as
well as spatially – depending upon the metric, to the extent that
flow of 'proper' time '$s$' gets reversed during half of
'coordinate' time '$t$'. In these solutions, time is ascribed the
topology $S^1/Z_2$ . Such space-times of Lorentzian signature can
be constructed by taking geometric product of this topology of
time, with 3-spaces, which could be the spatial section of
space-times such as, Minkowski, Schwarzschild. A new class of
geodesic closed curves, which are almost everywhere time-like
(except at critical reversal points), is obtained in the process.
Such curves may be termed as Closed Time-like Curves, almost
everywhere (CTCs,a.e.). We use the term recurrence metrics to
describe such space times, as, any initial distribution of
classical point particles, recurs after a period $2\pi$. Matter
tensors of some of these solutions are found to possess unusual
properties, such as periodic variation of pressure and density.
While this would be expected in periodic models, the physical
basis for such a variation is not clear. The paper therefore can
be regarded as an extension of Tipler's \cite {Tipler_1979} "no go
theorem for recurrence in an expanding universe", to other
space-time geometries.

\section{Recurrence metrics}
Consider a space-time in which time is having $S^1/Z_2$ topology.
This may be obtained from the Minkowski Universe as follows. First
identify the hypersurfaces $t=0$, and $t=2 \pi$. This assigns a
$S^1$ topology to time. Next identify end points of chords of
$S^1$, which makes topology of time $S^1/Z^2$. This is essentially
the set of chords of the circle. Line element for this space-time
- which we call the 'Periodic Minkowski' metric, is obtained from
Minkowski metric by the subsitution,
\begin{equation}\label{Sin t}
    t \rightarrow \sin t
\end{equation}
This gives the following line element,
\begin{equation}\label{PMink}
    ds^2=\cos^2 t dt^2-dr^2
\end{equation}
where,$r^2=x_1^2+x_2^2+x_3^2$, and $(x_1, x_2, x_3)$ are
rectangular co-ordinates.
\subsection{Consistency between Global Topology, and Geodesics}
Periodic topologies of time such as $S^1$ and $S^1/Z_2$, require
recurrence that Cauchy data at hypersurfaces $t$ and $t+2\pi$, be
identical. If $q_n(t)$ and $p_n(t)$ represent position and
momentum respectively, of $n$-th particle at time instant $t$,
then recurrence requirement implies,
\begin{equation}\label{}
    (q_n(t),p_n(t))=(q_n(t+2\pi),p_n(t+2\pi))
\end{equation}
for all $n$. Since non-interacting particles move along geodesics,
therefore, consistency between a compact topology of time and
geodesics, requires that all geodesics close in same period
$2\pi$. Further, as geodesics are derived from metric, via the
geodesic equation,
\begin{equation}\label{}
    \frac{d^2x^\mu}{ds^2}+\Gamma^\mu_{\rho
    \sigma}\frac{dx^\rho}{ds}\frac{dx^\sigma}{ds}
\end{equation}
the recurrence constraint has to be imposed through the metric. We
show in subsection on Geodesics Equations (below), that geodesics
of metric (\ref{PMink}), have a global geometry consistent with
the compact $S^1/Z_2$ time topology, i.e., all of them close in an
identical period $2\pi$. If on the other hand we transform metric
(\ref{PMink}) to,
\begin{equation}\label{Mink}
    ds^2=dt^2-dr^2
\end{equation}
and impose compact time topology by fiat - i.e., identifying the
hypersurfaces $t=0$, and $t=2\pi$, then the geodesic equations,
yield the following solution to particle motion,
\begin{equation}\label{}
    x_i(t)=C^1_i t+C^2_i
\end{equation}
for spatial coordinates $x_i$, $i=1,2,3$. Here $C^1_i$ and
$C^2_i$, can be interpreted as the initial velocity,  and the
initial position, respectively, of the particle. Its clear that
the geodesics of (\ref{Mink}) do not close. Thus the global
geometry of (\ref{Mink}) is inconsistent with the imposed periodic
topology of time. How do we resolve this conundrum? The answer
offered by author is that metrics (\ref{PMink}) and (\ref{Mink})
correspond to two different space-times, which we label $M_1$ and
$M_2$ respectively.  $M_2$ should be regarded as universal
covering space of $M_1$. Topology of time in $M_2$, should be
$R^1$, for it to be consistent with global geometry of its
geodesics. It should be kept in mind, that where as metric is a
local specification of space-time geometry, geodesics are global
objects, which relate to space-time's global topology. Thus while
locally $M_1$ and $M_2$ have identical geometries, however,
globally geometry of their geodesics differ.

\subsection{Proper time}
For a fixed value of $r$, we have for proper time $s$,
\begin{equation}\label{}
    s=\int \cos t dt
\end{equation}
Notice that proper time s is not a monotonically increasing
function, but periodically reverses direction. As coordinate time
$t$ varies monotonically between $[0,2\pi]$, proper time $s$
oscillates between $1$ and $-1$. Note that a system described at
time $t$, and undergoing time reversals at instants $t+\pi$ and
$t-\pi$ will also show recurrence. But here the system undergoes
discontinuous changes at moments of time reversals. On the other
hand in metric \ref{PMink}, the changes are smooth.

\subsection{Light cone}
For light cone we have following relation between r and t,
\begin{equation}\label{}
    \frac{dr}{dt}=\cos t.
\end{equation}
The behavior of light cone semi angle is same as that of the
function $\cos t$. The physical significance of such a time
varying light cone semi angle however has interesting implication.
$dr/dt$ decreases in the interval $[0, \pi]$, becoming zero at
$t=\pi/2$ -  indicating a gradual slowing down of all particles -
including light - with an instantaneous stop at $t=\pi/2$.
Further, dr/dt is negative in the interval $[\pi,3\pi/2]$,
indicating a reversal of direction of propagation of all
particles, and another stoppage at $t=\pi$. $dr/dt$ becomes
positive once again, in the interval $\pi/2,2\pi$, with implies
particles start moving in their initial direction, gradually
gaining speed.

\subsection{Loschimdt velocity reversion}
The points where dr/dt becomes zero are termed as critical points
in Morse theory [4], for functions defined on circles. It has been
pointed out, that one of the conditions for Poinc$\acute{a}$re
recurrence, is the Loschimdt velocity reversion, with velocity $v
\rightarrow -v$, simultaneously for all particles. As a result of
this velocity reversion, all particles would retrace their paths
backwards, leading to recurrence of initial position, but not the
initial velocity. For recurrence of initial velocities also, the
particles are allowed to go beyond their initial positions, and a
second velocity reversion is applied. This subsequently leads to
complete Poinc$\acute{a}$re recurrence – of both positions as well
as velocities. Such phenomena ordinarily do not occur in physical
situations, - but may be achieved by applying time reversal to the
system at instants $\pi$ and $\pi$. The smoothly time varying
light cones suggested here, allow this to happen smoothly. We
verify in sub-sections on Einstein's equations and Matter tensor
that metric in eq. (\ref{PMink}) is a solution of Einstein's field
equations. Below we verify periodicity of geodesic equations.

\subsection{Geodesic Equations}
The only non-zero Christoffel symbol is -
\begin{equation}\label{}
    \Gamma^0_{00}= - \tan t
\end{equation}
Substituting this in the geodesic equation gives following
equations for $t$ and $x_i$
\begin{eqnarray}
\nonumber t''(s)-t'(s) \tan t(s) =0,  \\
 x_i''(s)=0, i=1,2,3,
\end{eqnarray}

with solution of the form

\begin{eqnarray}
\nonumber t(s)=\arcsin [C^1_0 ( s-C^2_0)]  \\
\begin{array}{cc}
 x_i(s)=C^1_i s + C^2_i, & i=1,2,3.
 \end{array}
\end{eqnarray}
where, $C^1_\mu, and C^2_\mu, \mu=0,1,2,3$, are integration
constant. This yields the following relation for spatial
coordinate $x_i(t)$, and velocity $v_i(t)=x_i'(t)$, as periodic
function of time $t$.
\begin{eqnarray}\label{}
\nonumber   x_i(t)=\frac{C^1_i}{C^1_0}(\sin t+C^2_0) +C^2_i \\
    v_i(t)=\frac{C^1_i}{C^1_0}\cos t
\end{eqnarray}
Integration constants $C^1_\mu, and C^2_\mu$, can be interpreted
in terms of initial position $x{0}$ and initial velocity $v(0)$ as
follows -
\begin{eqnarray}
\nonumber
  x_i(0) = \frac{C^1_i C^2_0}{C^1_0}+C^2_i \\
  v_i(0) = \frac{C^1_i}{C^1_0}
\end{eqnarray}
which gives 8 unknown integration constants in terms of 6 initial
conditions. Setting
\begin{eqnarray}
\nonumber C^1_0=1 \\
C^2_0=0
\end{eqnarray}
gives $s=\sin t$ and allows determination of remaining integration
constants in terms of initial conditions.

\subsection{Einstein equations}
It can be verified that Riemann tensor vanishes,
\begin{equation}\label{}
    R_{\mu\nu\rho\sigma} = 0,
\end{equation}
and therefore Ricci tensor $R_{\mu\nu}$, Ricci scalar $R$, and
Einstein tensor $G_{\mu\nu}$ also vanish. Einstein equation with
cosmological constant $\Lambda=0$ is,
\begin{equation}\label{}
    G_{\mu\nu}=0=T_{\mu\nu},
\end{equation}
and for non-zero $\Lambda$ it is
\begin{equation}\label{}
   \Lambda \cos^2 t = T_{\mu\nu}
\end{equation}
where, $T_{\mu\nu}$ is the matter tensor. Following
interpretations for matter tensor are possible.

\subsubsection{$\Lambda=0$}
\begin{enumerate}

\item A vacuum solution with zero matter density $\rho$ and zero
pressure $p$. This however, is not interesting from recurrence
view point, as recurrence is supposedly for a system populated
with particles.

\item A solution with classical particles of zero mass $m=0$, and
therefore carrying zero momentum, and exerting zero pressure. Note
that quantum particles with $m=0$ such as photons will contribute
to universe's matter density, and therefore do not form a part of
this solution.

\item A solution with equal number of classical particles with
positive mass $m$, and negative mass $-m$. This would ensure zero
matter density at a macroscopic (coarse grained) scale. We note
that negative energy states in general relativity also occur in
other situations such as worm holes. Properties of negative mass
particles are discussed in \cite{Terletskii_1968}. Equal amounts
of Tachyonic matter of form $im$ and $-im$ would also be
admissible. Interaction between all these particles should be such
as to ensure zero pressure on macroscopic scale. Note that Dirac's
interpretation for the positive and negative energy solutions of
Klein-Gordon equation allows equal number of positive matter and
negative energy anti-matter. Whether universe is matter-antimatter
asymmetric or not has been a source of debate. It has been
suggested that one of possible origins for high energy cosmic rays
is matter-antimatter annihilation occurring in the universe.

\end{enumerate}

\subsubsection{$\Lambda \neq 0$}
This yields,
\begin{eqnarray}\label{Lam1}
 \nonumber \rho=\Lambda (\cos^2 t+1) \\
p=-\Lambda
\end{eqnarray}
So either the pressure is negative or density is negative -
depending upon the sign of $\Lambda$.

From above possibilities for matter tensor, we see that there are
severe constraints on matter states and inter-particle interaction
\footnote{Sign of pressure depends upon whether inter-particle
forces are attractive or repulsive.} for recurrence in general
relativity. Behavior of density in particular in eqs. (\ref{Lam1})
needs physical explanation. A qualitative attempt for this is made
in section on "Speculation on periodic mass acquisition" in
particle models.

\subsection{Wave equation}
The wave equation for $1-D$ form of (\ref{PMink}) is
\begin{equation}\label{}
    \frac{1}{\cos^2 t}\frac{\partial^2 \phi}{\partial t^2}
    -\frac{\partial^2 \phi}{\partial x^2} = 0
\end{equation}
with the solution
\begin{equation}\label{AdWave}
    \phi = \exp^{[kx-\omega \sin t]}+\exp^{-[kx-\omega \sin t]}
\end{equation}
where the first term represents the retarded component and the
second term represents the advanced component, of the solution. It
can be seen, that advanced wave differs from retarded wave only by
a phase $\pi$. Usually the advanced waves are considered as time
reversed retarded waves. However, in the (\ref{AdWave}), advanced
waves can be interpreted as retarded waves, returning to a point
after a time interval of $\pi$.

\subsection{Causality violation}
There exist a number of solutions of  general relativity having
Closed Time-like Curves (CTCs).  G\"{o}del \cite{Godel_1949} was
the first one in which existence of CTCs was demonstrated.  The
metric presented here, generates a closed curve in space-time,
which consists of two time-like segments joined at ends. However,
this curve is space-like at points of join. The closed curve thus
generated can hence be termed as a "CTC, almost everywhere" (CTC,
a.e.).

\subsection{Background fluctuation}
Random background fluctuations will destroy recurrence. However,
fluctuations with time periodic constraint, (i.e., which can be
expanded as a commensurate Fourier series) will preserve
recurrence.

\section{Periodic Schwarzschild type universes}
The maximal extension of Schwarzschild line element, in a form
without coordinate singularity \cite{Carmeli_1982} is,
\begin{equation}\label{Sch}
    ds^2=f^2 dt^2-f^2 dr^2-r^2d\theta^2-r^2\sin^2
    \theta d\phi^2
\end{equation}
where,
\begin{equation}\label{}
    f^2=\frac{32Gm^3}{r} \exp \left(-\frac{r}{r_s} \right)
\end{equation}
Consider the modified Schwarzschild line element in which $t$ is
compactified to $S^1/Z_2$ topology, as in the periodic Minkowski
case. The line element is now obtained from eq.(\ref{Sch}) by the
using the substitution of eq.(\ref{Sin t}), which gives,
\begin{equation}\label{PtSch}
ds^2=f^2\cos^2 tdt^2
    -f^2r^2-r^2d\theta^2-r^2\sin^2\theta d\phi^2
\end{equation}
Here $R_s$  is the Schwarzschild radius. It can be verified that
eq.(\ref{PtSch})is a Ricci flat metric and therefore constitutes a
vacuum solution of Einstein's field equations. Light cone semi
angle is  same as in the Periodic Minkowski case. Space-time is
now bounded in past ($t=-\pi /2$) and future ($t=\pi /2$), which
are points of reversion. $g_{00}=0$ while $g_{11}$ remains finite
at these points of reversion. Space-time reduces to 3 spatial
dimensions at these points. $g_{00}$ and $g_{11}$ are finite at
$r=r_s$.  At $r=0$ and $t=\pm \pi /2$, $g_{00}=32Gm^3$, is finite,
while $g_{11}$ is infinite.

This metric, however does not allows radiation and massive
particles to escape to infinity because of periodic time
reversals. Similarly particles with initial position $r_0$ and
radially inward velocity $v(t)$ satisfying
\begin{equation}\label{}
    r_0 > 2 \int_0^1 v(t) dt
\end{equation}
will not reach the Schwarzschild radius and continue to oscillate
outside the horizon, due to time reversals. The limits  of above
integral correspond to the interval in which $t$ is positive,
i.e., $t \in [0,1]$. During the time reversed phase, the black
hole will behave as a white hole, as the particles would be coming
out of the horizon. They however will not escape to infinity
(again because of time reversal), and will oscillate around the
horizon.

It is expected that our universe is governed by a uniform arrow of
time. Can space-time (\ref{PtSch}),  with a locally oscillating
time arrow, be embedded in a universe, with asymptotically,
uniform, time arrow? The answer is provided by the following
substitution for time co-ordinate $t$, outside horizon,
\begin{equation}\label{tExp}
    t \rightarrow t + e^{[r_s-r] (\sin t -t)}
\end{equation}
This choice of $t$ gives a behavior as $\sin t$, near the horizon,
and the usual $t$ behaviour for $r>>r_s$ . A further
transformation
\begin{equation}\label{}
    t \rightarrow t^\gamma
\end{equation}
allows periodicity to increase or decrease with time, depending
upon sign of $\gamma$.

\section{Discussion}
Periodic Minkowski is singularity free and has a admissible vacuum
solution. It also admits a solution with a universe having an
equal number of positive and negative masses, which however is
un-physical. The solution of Periodic Minkowski with non-zero
density has un-physical periodic variation in density. Periodic
Schwarzschild type metric admits a vacuum solution. Present work
can be regarded as a continuation of exploration by Tipler
\cite{Tipler_1979}.

\end{document}